\documentclass{ws-procs9x6}
\usepackage{graphicx}

\begin{document}

\title{Breathers in FPU systems, near and far from the
phonon band \footnote{ \uppercase{P}roceeding of the conference on
{\em ``\uppercase{L}OCALIZATION AND ENERGY TRANSFER IN NONLINEAR
SYSTEMS"}, \uppercase{J}une 17-21, 2002, \uppercase{S}an
\uppercase{L}orenzo de \uppercase{E}l \uppercase{E}scorial,
\uppercase{M}adrid, \uppercase{S}pain. \uppercase{T}o be published
by \uppercase{W}orld \uppercase{S}cientific.}}

\author{B. S\'anchez-Rey\dag, JFR. Archilla\dag, G
James\ddag, and J. Cuevas\dag}
\address{\dag Nonlinear Physics Group, University of
Sevilla, Spain\\ \ddag D\'epartement de G\'enie Math\'ematique,
INSA de Toulouse, France \\Email:bernardo@us.es} \maketitle

{\bf Introduction.} This work is motivated by a recent breathers
existence proof in the one dimensional FPU system, given by the
equations:
\begin{equation}
\label{eq:fpu} \ddot{x}_n = V'(x_{n+1}-x_n)-V'(x_n - x_{n-1})\; ,
\quad n\in \mathbb{Z} \;,
\end{equation}
where $V$ is a smooth interaction potential satisfying
$V^{\prime}(0)=0$ and $V^{\prime\prime}(0)>0$. Using a center
manifold technique\cite{J01}, one can prove the existence of small
amplitude breathers (SAB) with frequencies $\omega_{b}$ slightly
above the phonon band if $B=\frac{1}{2} V''(0)
V^{(4)}(0)-(V^{(3)}(0))^{2}>0$, and their non-existence for $B<0$.
Our aim is to test numerically the range of validity of this
theoretical result and to explore new phenomena. For this purpose
we shall fix $V(u)=u^2/2+a \,u^3 +\frac{1}{4}\,u^4$, which yields
$B=3(1-12a^{2})$.

     We work with the difference variables
$u_n=x_n-x_{n-1}$ more suitable for the use of our numerical
method. We also use periodic boundary conditions
$u_{n+2p}(t)=u_{n}(t)$ so that the maximum frequency of the linear
phonons is exactly 2 as in the infinite lattice. Our computations
are performed using a numerical scheme based on the
anti-continuous limit and Newton method\cite{MA}.

\begin{figure}
\begin{center}
\begin{tabular}{cc}
\includegraphics[width=5.5cm]{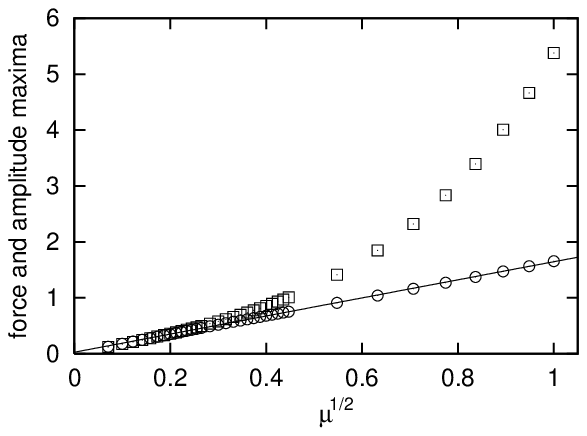}&
 \includegraphics[width=5.5cm]{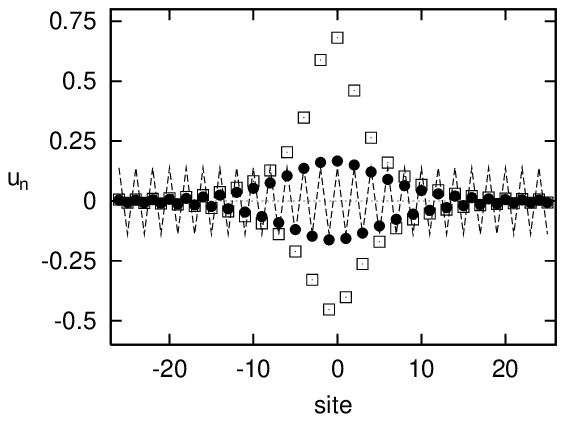}
\end{tabular}
\end{center}
\caption{Left: Force (squares) and amplitudes (circles) maxima
versus $\mu^{1/2}$. The cubic coefficient in $V$ is $a=-0.1$
($B=2.64$). Right: Comparison between a SAB (full circles) for
$a=-0.1$ and a LAB (blank squares) for $a=-1/3$ ($B=-1$) having
the same frequency $w_b=2.01$. The dashed line represents the
linear phonon with frequency~$2$. \label{fig1}}
\end{figure}

{\bf Test and range of validity.}   First, we have computed
numerically SAB (i.e. breathers whose amplitudes go to zero when
$w_b \rightarrow 2^{+}$) in the case  $B>0$. We have obtained
breathers with symmetries $u_n(t)=u_{-n}(t)$ (Page mode) and
$u_n(t)=u_{-n-1} (t+T_{b}/2)$ (Sievers-Takeno mode), where
$T_{b}=2\pi/\omega_{b}$ is the breather period.  The force
$y_n=V'(u_n)$ is the variable used in reference \cite{J01}. In
Fig.\ref{fig1} (left) it is shown that the maxima of the force are
of order $\mu^{1/2}$ when $\mu=w_b-2\rightarrow 0^{+}$, as
predicted by the theory, up to relatively large values.  Thus if
$B>0$ breathers exist for any small value of energy in our FPU
system (\ref{eq:fpu}).

     Another property of these SAB  is that their width diverges when
$w_b \rightarrow 2^{+}$. More precisely the theory predicts that
their spatial extend is of order $\mu^{-1/2}$, which is in
accordance with our numerical observations.

{\bf Other numerical observations.} For $B>0$, we have numerically
continued the SAB as $\omega_{b}$ goes away from the phonon band.
We have found that the maximum amplitudes of the oscillations
are also approximately linear functions of
$\mu^{1/2}$. This is expected for small $\mu$, since
$u_{n}=y_{n}+O(y_{n}^{2})$, but it occurs surprisingly far from
the phonon band, at least until values of $\mu \approx 1$ (see
fig.\ref{fig1}, left). We have also checked that the Page mode
fits very well to the NLS soliton $u_n(t)=\alpha\, \sqrt{\mu }\,
(-1)^{n}\, \cos{(\omega_{b} t)}\, [ \cosh{(\beta\, \sqrt{\mu }\,
n)} ]^{-1}$, even far from the top of the phonon band.

\vspace{-0.25cm}

\vspace{-0.3cm}
    For $B<0$ and $V$ strictly convex
($\frac{1}{\sqrt{12}}<|a|<\frac{1}{\sqrt{3}}$), breathers exist
near the top of the phonon band but they are large amplitude
breathers\cite{AKK01} (LAB), i.e. their amplitudes do not go to
zero when $w_b \rightarrow 2^{+}$.  As a consequence there is an
energy gap for breathers creation in these FPU systems. In figure
\ref{fig1} (right) we compare a SAB and a LAB having the same
frequency $w_b=2.01$. We have found LAB with the same symmetries
as SAB (Page and Sievers-Takeno modes). The Page mode fits very
well to an exponential profile having the form $u_n(t)=\alpha
(\omega_b)\, (-1)^{n} \cos{(\omega_{b} t)} \, |\sigma (\omega_b
)|^{|n|}$ where $\sigma (\omega_b )= 1-(\omega_b^2)/2+
(\omega_b/2) (\omega_b^2-4)^{1/2} \in (-1,0)$. As (\ref{eq:fpu})
is formulated as a mapping in a loop space\cite{J01} and
$\omega_{b}>2$, the linearized operator has a purely hyperbolic
spectrum and the constant $\sigma(\omega_b ) $ is the closest
eigenvalue to $-1$ (with $\sigma (2)=-1$). Consequently, for
$\omega_{b}\approx 2$ one can ask if the iterated map admits a
global center manifold containing these LAB.

\end{document}